# Topological Electronic Structure and Its Temperature Evolution in Antiferromagnetic Topological Insulator MnBi$_2$Te$_4$

Y. J. Chen[1*], L. X. Xu[2,3,4*], J. H. Li[1], Y. W. Li[5], C. F. Zhang[6], H. Li[7,8], Y. Wu[1,8], A. J. Liang[2,9], C. Chen[2,9], S. W. Jung[10], C. Cacho[10], H. Y. Wang[2,4], Y. H. Mao[6], S. Liu[2], M. X. Wang[2], Y. F. Guo[2], Y. Xu[1,11,12], Z. K. Liu[2†], L. X. Yang[1,11†], and Y. L. Chen[1,2,5†]

[1]*State Key Laboratory of Low Dimensional Quantum Physics, Department of Physics, Tsinghua University, Beijing 100084, P. R. China*
[2]*School of Physical Science and Technology, ShanghaiTech University and CAS-Shanghai Science Research Center, Shanghai 201210, China*
[3]*Center for Excellence in Superconducting Electronics, State Key Laboratory of Functional Material for Informatics, Shanghai Institute of Microsystem and Information Technology, Chinese Academy of Sciences, Shanghai 200050, China*
[4]*University of Chinese Academy of Sciences, Beijing 100049, China*
[5]*Department of Physics, Clarendon Laboratory, University of Oxford Parks Road, Oxford, OX1 3PU, UK*
[6]*College of Advanced Interdisciplinary Studies, National University of Defense Technology, Changsha 410073, China*
[7]*School of Materials Science and Engineering, Tsinghua University, Beijing, 10084, China*
[8]*Department of Physics and Tsinghua-Foxconn Nanotechnology Research Center, Tsinghua University, Beijing, 100084, China*
[9]*Advanced Light Source, Lawrence Berkeley National Laboratory, Berkeley, California 94720, USA*
[10]*Diamond Light Source, Harwell Campus, Didcot OX11 0DE, UK*
[11]*Collaborative Innovation Center of Quantum Matter, Beijing 100084, P. R. China*
[12]*RIKEN Center for Emergent Matter Science (CEMS), Wako, Saitama 351-0198, Japan*

*These authors contributed equally to this work.
†*Email address:* [liuzhk@shanghaitech.edu.cn](liuzhk@shanghaitech.edu.cn), [lxyang@tsinghua.edu.cn](lxyang@tsinghua.edu.cn), [yulin.chen@physics.ox.ac.uk](yulin.chen@physics.ox.ac.uk)

**Topological quantum materials coupled with magnetism can provide a platform for realizing rich exotic physical phenomena, including quantum anomalous Hall effect [1-4], axion electrodynamics [5-7] and Majorana fermions [8]. However, these unusual effects typically require extreme experimental conditions such as ultralow temperature [2, 4, 9] or sophisticate material growth and fabrication [5, 7]. Recently, new intrinsic magnetic topological insulators were proposed in $MnBi_2Te_4$-family compounds – on which rich topological effects [10-12] could be realized under much relaxed experimental conditions [13, 14]. However, despite the exciting progresses, the detailed electronic structures observed in this family of compounds remain controversial [15-20] up to date. Here, combining the use of synchrotron and laser light sources, we carried out comprehensive and high resolution angle-resolved photoemission spectroscopy studies on $MnBi_2Te_4$, and clearly identified its topological electronic structures including the characteristic gapless topological surface states. In addition, the temperature evolution of the energy bands clearly reveals their interplay with the magnetic phase transition by showing interesting differences for the bulk and surface states, respectively. The identification of the detailed electronic structures of $MnBi_2Te_4$ will not only help understand its exotic properties, but also pave the way for the design and realization of novel phenomena and applications.**

Topological quantum materials (TQMs) represent special classes of materials whose electronic structures can be characterized by topological invariants protected by certain symmetries, and the breaking of these symmetries can lead to intriguing topological phase transitions. Time reversal symmetry (TRS), for instance, is such an important symmetry in TQMs. While the TRS can protect the unique helical gapless surface states in topological insulators (TIs), it is broken in many exotic but interesting topological phases, such as quantum anomalous Hall (QAH) insulators [1-4], topological axion insulators [5-7], magnetic Dirac and Weyl semimetals [21-26] and compounds showing Majorana fermions [8]. However, despite the many theoretical proposals [1, 21-27], finding real TQMs with spontaneous broken TRS is challenging. For example, up to date, QAH effect could only be realized in TI $(Bi_{1-x}Sb_x)_2Te_3$ thin films with fine-tuned magnetic doping (e.g. by V- or Cr-doping), under a temperature much lower than their Currie temperature [2, 4]. Therefore, it is highly urged to search for stoichiometric TQMs with intrinsic magnetic structures for the realization of high-temperature QAH effect and other exotic topological properties.

Recently, it was realized that a layered compound, $MnBi_2Te_4$ could be a promising magnetic TI with intrinsic A-type anti-ferromagnetic (AFM) order [10-12] that can serve as an ideal platform for the realization of QAH effects and other interesting magnetic topological phases including axion insulator and ideal Weyl semimetal phases [11]. Indeed, experimental efforts soon followed and topological phenomena such as magnetic-field-induced QAH effect [13] and quantum phase transition from axion insulator to Chern insulator [14] were recently observed. In order to understand the rich properties and explore the full application potential of $MnBi_2Te_4$, both the surface and bulk electronics structures need to be experimentally verified.

Unfortunately, although there have been several attempts to resolve the band structures of this interesting compound, the results remain controversial and even conflicting [15-20]. First of all, the identification of the topological surface states (TSSs) remains elusive, leading to significant difference in identifying the band gap of the surface state (due to the formation of magnetic order), ranging from 50 meV [15] to a couple of hundreds of meVs [16-19]. More intriguingly, these assumed gaps of the topological surface state persist well above the Néel temperature ($T_N$) of MnBi$_2$Te$_4$ (25K, see ref [17]) and could even be observed at room temperature [16, 18]. The order of magnitude's difference between the room temperature (300K) and the $T_N$ of MnBi$_2$Te$_4$ (25K) make it less convincing that the observed gap originates from the magnetic order. Finally, the fact that the measured gap size varies with photon energy [15-19], further alludes its bulk origin. These puzzles thus demand a systematic investigation on the electronic structure of MnBi$_2$Te$_4$ that can clearly distinguish the surface and bulk states in order to reveal their connection to the AFM transition – which can lay the foundation to the understanding of the interplay between the electronic structure, magnetic ordering and the rich topological phenomena of this material.

In this work, combining the use of synchrotron and laser light sources, we carried out comprehensive and high resolution angle-resolved photoemission spectroscopy (ARPES) studies on MnBi$_2$Te$_4$, and clearly identified its topological electronic structures including the characteristic TSSs. The broad photon energy range provided by the synchrotron light source allowed us to map out the subtle three-dimensional (3D) electronic structures; while the high resolution laser light source made it possible to investigate the TSSs with great precision. Furthermore, by carrying out temperature dependent measurements, we were able to observe the temperature evolution from both the bulk and surface state bands, which shows interesting difference between their interplay with the magnetic phase transition: while the bulk states show clear splitting commencing at $T_N$; the gap of TSSs is negligible within our energy resolution [2 mV at temperature (7.5 K) well below $T_N$], which could be due to the highly delocalized nature of TSSs mediated by surface magnetic domains of different magnetization orientations. The identification of the detailed electronic structures of MnBi$_2$Te$_4$ will not only help understand its exotic properties, but also pave the way for the design and realization of novel phenomena and applications.

MnBi$_2$Te$_4$ crystallizes into a rhombohedral lattice with space group of $R\bar{3}m$ [19, 28]. It exhibits a layered structure by staking van der Waals septuple Te-Bi-Te-Mn-Te-Bi-Te layers, which has an extra Mn-Te layer sandwiched at the middle of the well-known quintuple-layer of $Bi_2Te_3$, as shown in Fig. 1a. The high quality of the single crystalline samples used in this work is demonstrated by the single crystal X-ray diffraction patterns (Fig. 1b) and angle scan (Fig 1c). The magnetic susceptibility and electric transport measurements (Fig. 1d) clearly show the AFM phase transition at $T_N = 25$ K. For the ARPES measurements, the sample was cleaved in-situ with flat surfaces, showing clear layered structure (Fig. 1b). Scanning tunnelling microscopy measurement reveals step height of about 1.37 nm on the sample surface, in good consistence with the thickness of the septuple layer (1/3 $c$) in $MnBi_2Te_4$. The characteristic Mn, Bi and Te core level peaks in X-ray photoemission spectroscopy measurement further confirm the sample phase (Fig. 1f). The broad band structure mapping at the surface Dirac point [~0.27 eV below the Fermi energy ($E_F$), as we will discuss later] over multiple Brillouin zones (BZs) shows clear point-like features with hexagonal symmetry (Fig. 1g), in consistence with the crystal structures of $MnBi_2Te_4$.

The bulk band structure of $MnBi_2Te_4$ is illustrated in Fig. 2. The conduction and valance bands form an inverted bulk gap [10, 11] whose magnitude varies with the photon energies used for the ARPES measurements (Fig. 2a, b), indicting its bulk nature. Depending on the methods used for gap extraction (see SI for details), the bulk gap size at different $k_z$ momentum varies from 180~220meV [fitted peak-peak gap in energy distribution curves (EDCs)] or 80~160meV (EDC leading edge gap), as summarized in Fig. 2b(iii). Interestingly, besides the bulk bands with very strong photoemission spectra intensity that agree well with the calculations (Fig. 2c(i)), one can notice that there is also weak spectra intensity within the bulk band gap (Fig. 2c(ii-iv)), showing an "X" shape dispersion centered at the Γ point, similar to the dispersion of the TSSs of other 3D TIs [29, 30].

To further investigate the TSSs in the bulk band gap, we carried out laser based ARPES measurements with superb energy and momentum resolution, as illustrated in Fig. 3. The evolution of band structures with different binding energy (Fig. 3a, b) clearly shows the gapless TSSs with conical shape dispersion across the Dirac point located at $E \sim -0.27$ eV within the bulk band gap. The linear dispersion of these TSSs can be directly seen in Fig. 3c, with vanishing energy gap less than the energy resolution of the experiments ($\Delta E$~2 meV), which can be better seen in the zoom-in plot around the Dirac point (Fig. 3d).

Besides the TSSs, the high resolution in the measurement of laser based ARPES also allow us to investigate the bulk states with great details. From the spectra intensity map (Fig.3c), the 2$^{nd}$ derivative plot (Fig. 3e) or the stacking momentum distribution curves (MDC) plot (Fig. 3f), one can clearly see three conduction bands. Interestingly, by comparing with theoretical calculation of bulk bands in Fig. 2c(i), we notice that an extra bulk band is observed, whose origin will be discussed later. We mark the observed conduction bands as CB2, CB1$_a$, and CB1$_b$ respectively as shown in Fig. 3e, f (the reason of the naming will be discussed below).

To understand the interplay between the band structure and the magnetic properties of MnBi$_2$Te$_4$, we carried out a series of temperature dependent ARPES measurements across $T_N$, which are summarized in Fig. 4. For the bulk conduction bands, upon the increase of temperature, the CB1$_a$ and CB1$_b$ bands move towards each other and eventually merge into one single band (CB1) above $T_N$, as illustrated in Fig. 4a, b. When the temperature is sequentially cooled down to below $T_N$, CB1 splits into CB1$_a$ and CB1$_b$ again, as can be clearly seen in Fig. 4a(vi) and Fig.4b(vi), the same behavior has been observed in the measurements on multiple samples (see SI for more details). The side-by-side comparison between the band structure above and below $T_N$ can be seen in Fig. 4c for clarity and the temperature evolution of the EDCs at the Γ point is summarized in Fig. 4d, with the fitted peak-to-peak splitting of the CB1$_a$ and CB1$_b$ on multiple temperatures plotted in Fig. 4e. The coincidence of the splitting temperature and the $T_N$ near 25 K indicates a correlation between the band splitting and the formation of the magnetic order. Although the observed bulk band splitting is beyond expectation of an AFM system, it is accessible considering the surface ferromagnetic ordering, which has been shown to induce considerable exchange splitting in AFM EuRh$_2$Si$_2$ [31]. On the other hand, according to first-principles calculations, CB1$_a$ and CB1$_b$ are mainly contributed by Te $p_z$ orbital. When the temperature increases above $T_N$, the exchange induced splitting vanishes, which has complicated influence on interlayer coupling and band dispersion, leading to the merge of the $CB1_a$ and $CB1_b$ bands.

Interestingly, in contrast to the bulk bands, the TSSs show no observable temperature dependence (Fig. 4a, b), with diminishing gap in a large temperature range. In the AFM MnBi$_2$Te$_4$, the TSSs can be described by an effective Hamiltonian $H(\boldsymbol{k}) = \sigma_x k_y - \sigma_y k_x + \tilde{m}_z \sigma_z$, where $\sigma_{x,y,z}$ are the Pauli matrices for spins, and $\tilde{m}_z$ is the effective mass induced by exchange interactions on the surface, which could open a surface gap of $2|\tilde{m}_z|$. For $T > T_N$, the system becomes paramagnetic, TRS gets preserved, and thus MnBi$_2$Te$_4$ transits into the time-reversal invariant TI phase, implying a vanishing surface gap. When the temperature is lowered below $T_N$, the material will transit into A-type AFM state, as revealed by previous studies [10, 11]. Therefore, if the surface magnetism is ideally ordered along the out-of-plane $z$ direction, a surface gap of ~40 meV would develop, as indicated by theoretical calculations [10]. The observed surface gap, however, is vanishingly small, suggesting that the surface magnetism may not be well ordered. The result could presumably be explained by following mechanisms. Firstly, the interlayer antiferromagnetic interactions are weaker on the surface than in the bulk, and the magnetic dipole-dipole interactions of ferromagnetic monolayer typically prefers an in-plane easy axis, which makes the orientation of surface magnetism easily fluctuated. The surface magnetism, once oriented along the in-plane $x$ direction, would contribute an interaction term of $\tilde{m}_x \sigma_x$ into the effective Hamiltonian, which cannot open the surface gap as confirmed by recent calculations [32]. This would shift the Dirac point of TSSs slightly away from the $\Gamma$ point. However, the averaged shift over different in-plane orientations would vanish due to the high symmetry of the in-plane structure. Secondly, magnetic domains ubiquitously exist in AFM materials and cannot be easily eliminated even by field cooling. Since the sign of exchange-induced mass depends on the orientation of surface magnetism, opposing magnetic domains would lead to opposite exchange effects. It is well known that Dirac-like surface states can have unusually large localization length $\lambda_D(T)$ caused by the Berry phase of $\pi$ [33] and

$\lambda_D(T)$ greatly enhances with decreasing temperature. At low temperatures $\lambda_D$ is expected to significantly exceed the size of magnetic domains (typically on the order of 10 nm) [34]. Thus, the effective exchange effects experienced by TSSs are contributed by a few magnetic domains, whose averaged effects are weakened by magnetic compensation. From another point of view, gapless chiral boundary modes are topologically protected to exist in the presence of opposing magnetic domains [35]. The gapless modes would play an important role in determining macroscopic properties of TSSs if the domain size is small enough (e.g., compared to $\lambda_D(T)$). This could also help explain why no magnetization-induced surface gap observed in MnBi$_2$Te$_4$.

In conclusion, we have presented a systematic investigation of the electronic structure of AFM-TI candidate MBT. We observed gapped bulk electronic bands with clear $k_z$ dispersion and a gapless TSS as the characteristic topological electronic structure which is attributed to the neutralized net magnetic moment of domains with different magnetization directions. We further observed bulk band splitting possibly related to the surface ferromagnetism caused by the Mn atoms buried in the surface layer. Our results not only reconcile the contradicting results on the origin and magnitude of previously observed gap in MBT, but also provide important insights into the electronic structure and magnetism of MBT, which is crucial for the understanding of the interplay between magnetism and topology in AFM TIs.

*Note Added*: During preparing this manuscript, another work reported gapless surface electronic states on MnBi$_2$Te$_4$ single crystals, which is in good agreement with our observations [36].

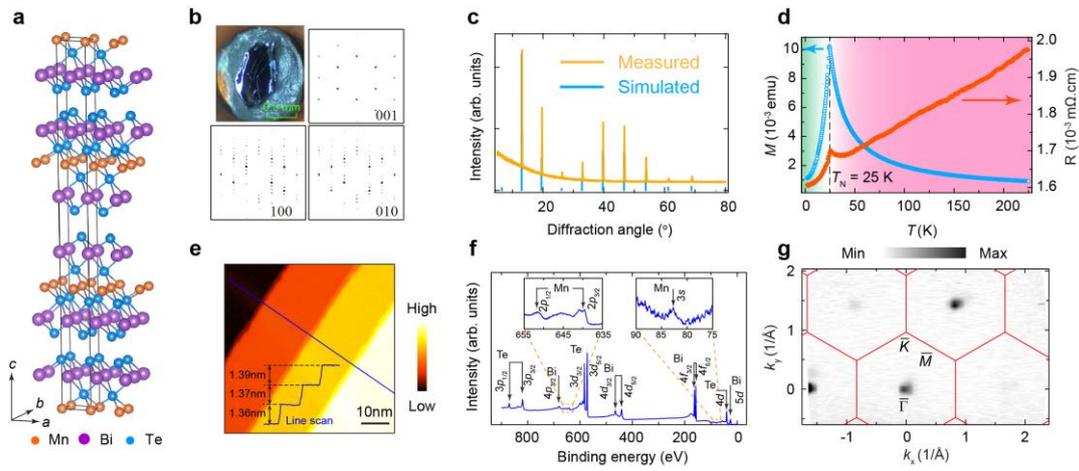

**Figure 1 | Basic properties and characterization of MnBi$_2$Te$_4$ single crystals. a** Schematic illustration of the layered crystal structure of MnBi$_2$Te$_4$. **b** Image of MnBi$_2$Te$_4$ single crystal after cleavage and X-ray diffraction patterns along different directions. **c** Angle-scan of X-ray diffraction along c-axis. **d** Magnetization (with magnetic field applied along *c* axis) and resistance as functions of temperature. **e** Scanning tunneling microscopy mapping of surface topography obtained with I = 250 mA and V = 1 V. **f** X-ray photoemission spectroscopy showing characteristic Mn, Bi and Te core levels. (g) Constant energy contour near 270 meV below the Fermi energy over multiple BZ.

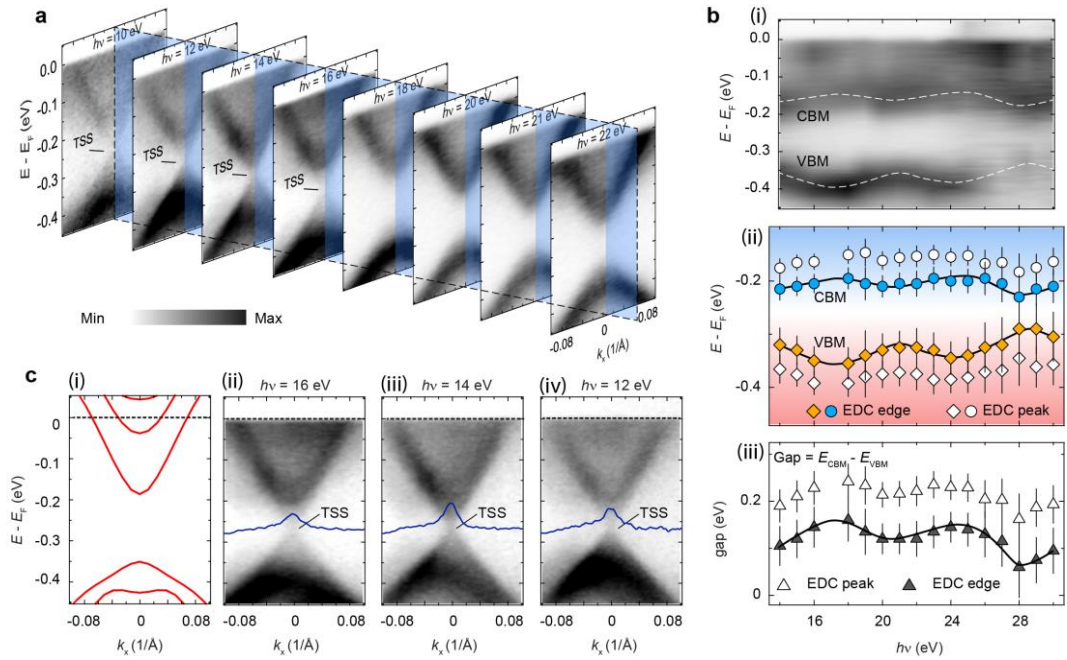

**Figure 2 | Photon energy dependent band structure measurements on MnBi$_2$Te$_4$. a** Band dispersion near $\bar{\Gamma}$ measured with different photon energies. The TSS as marked is observed at photon energies lower than 16 eV. **b** Photon energy dependence of (i) EDC near $\bar{\Gamma}$, (ii) extracted conduction band minimum (CBM) and valance band maximum (VBM) and (iii) bulk band gap. **c** (i) Calculated bulk band dispersion. (ii-iv) Band dispersion obtained at low-photon energies showing topological surface states (TSSs). The insets are the momentum distribution curves (MDCs) integrated near the Dirac point.

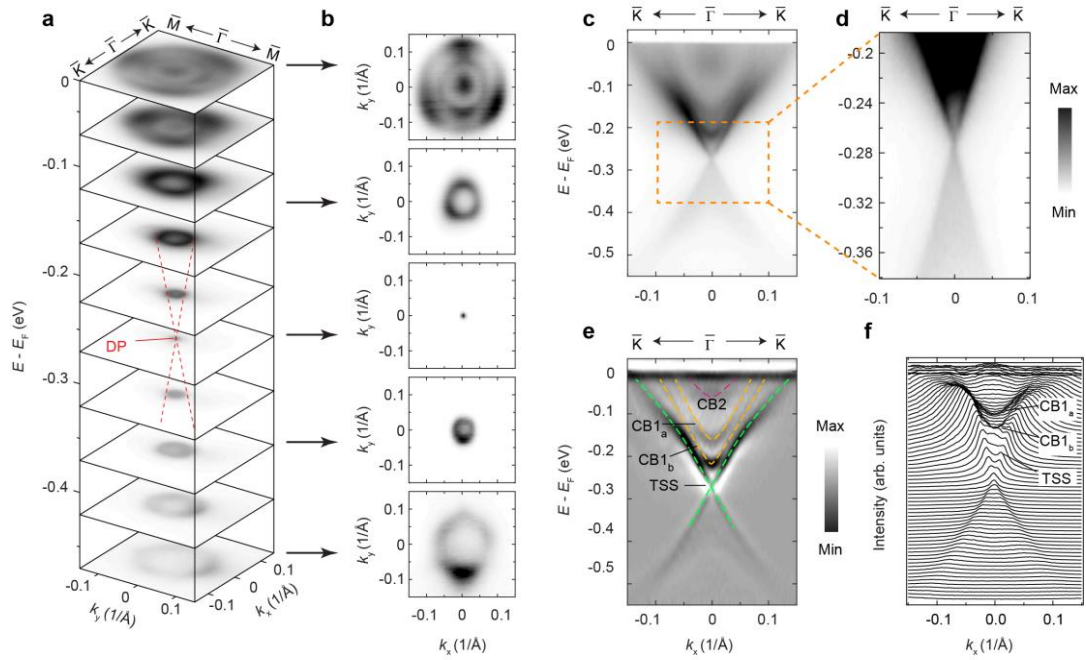

**Figure 3 | Electronic structure of MnBi$_2$Te$_4$ measured with 7 eV laser**. **a, b** constant energy contours at different binding energies. **c** Band dispersion along the $\overline{\Gamma K}$ direction with a clear gapless Dirac cone formed by topological surface state. **d** Zoom-in plot of the dispersion near the Dirac point. **e** Second derivative of ARPES intensity map in panel **c**. The dashed curves are guides-to-eyes for band dispersions. **f** Stacking plot of momentum distribution curves. The bulk conduction bands and gapless TSS are clearly observed. Data were collected at 7.5 K.

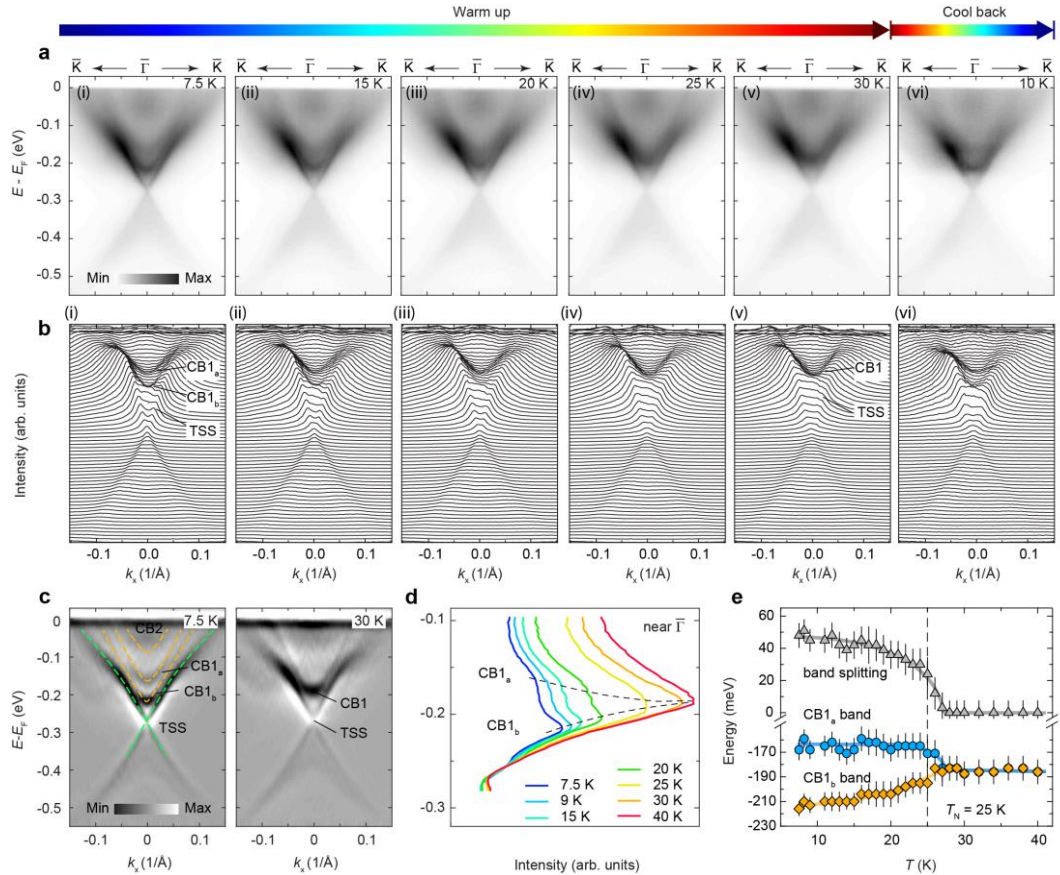

**Figure 4 | Temperature evolution of the band structure of MnBi$_2$Te$_4$. a, b** Band dispersions along $\bar{\Gamma}\bar{K}$ (**a**) and corresponding MDCs (**b**) at selected temperatures. **c** Side-by-side comparison of the second derivative of ARPES intensity maps at 7.5 K and 30 K. **d** Temperature evolution of EDCs at $\bar{\Gamma}$. **e** The energy position of CB1$_a$ and CB1$_b$ bands together with the energy difference between them as functions of the temperature showing a band splitting below 27 K. Data were taken with 7 eV laser.

**Methods:**

**Crystal growth:**

The single crystal of MnBi$_2$Te$_4$ was grown by a direct reaction of a stoichiometric mixture of Bi$_2$Te$_3$ and MnTe, which were synthesized by reacting high-purity Bi (99.99%, Adamas) and Te (99.999%, Aladdin), and Mn (99.95%, Alfa Aesar) and Te (99.999%, Aladdin), respectively. The growth of high-quality MnBi$_2$Te$_4$ single crystals was carried out in a sealed silica ampoule under a dynamic vacuum. The ampoule was firstly heated to 973 K. After being slowly cooled down to 864 K, the crystal growth occurred during the long-term annealing at the same temperature and afforded mm-sized shiny single crystals of MnBi$_2$Te$_4$. The crystals were examined on a PANalytical Empyrean diffractometer with Cu Kα radiation to establish the correct phase and high-quality of single crystal samples for the experiment.

**Angle-resolved photoemission spectroscopy:**

ARPES measurements were performed at beamline I05 of the Diamond Light Source (DLS), UK, beamline 13U of National Synchrotron Radiation Laboratory (NSRL) in Hefei, China, and beamline 10.0.1 of the Advanced Light Source (ALS), USA. The samples were cleaved *in-situ* and measured under ultra-high vacuum below $1 \times 10^{-10}$ Torr at DLS, $6 \times 10^{-11}$ Torr at NSRL and $3 \times 10^{-11}$ Torr at ALS. Data were collected by Scienta R4000, DA30 L and R4000 analysers at DLS, NSRL and ALS respectively. The total convolved energy and angle resolutions were 15 meV and 0.2°, respectively.

High-resolution laser-based ARPES measurement were performed at home-built setups (hv=6.994eV) in Tsinghua University and ShanghaiTech University. The samples were cleaved *in-situ* and measured under ultra-high vacuum below $6 \times 10^{-11}$ Torr. Data were collected by a DA30 L analyser. The total convolved energy and angle resolutions were 2 meV and 0.1°, respectively.

**First-principles calculations:**

First-principles calculations were performed by density functional theory (DFT) using the Vienna ab initio Simulation Package . The plane-wave basis with an energy cutoff of 350 eV was adopted. The electron-ion interactions were modeled by the projector augmented wave potential and the exchange-correlation functional was approximated by the Perdew-Burke-Ernzerhof type generalized gradient approximation (GGA) [37]. The GGA+U method was applied to describe the localized 3d-orbitals of Mn atoms, for which U = 4.0 eV was selected according to our previous tests [10]. The structural relaxation for optimized lattice constants and atomic positions was performed with a force criterion of 0.01 eV/ Å and by using the DFT-D3 method to include van der Waals corrections. Spin-orbit coupling was included in self-consistent calculations and the Monkhorst-Pack k-point mesh of $9\times9\times3$ was adopted. Surface state calculations were performed with WannierTools package [38], based on the tight-binding Hamiltonians constructed from maximally localized Wannier functions (MLWF) [39] .

**Acknowledgment:**

We thank Jinsong Zhang, Ke He, Zhong Wang and Yayu Wang for inspiring discussions. This work was supported by the National Natural Science Foundation of China (Grant No. 11774190, No. 11674229, No. 11634009, and No. 11774427), the National Key R&D program of China (Grant No. 2017YFA0304600 and 2017YFA0305400), EPSRC Platform Grant (Grant No. EP/M020517/1). Y. F. G. acknowledges the support from the Shanghai Pujiang Program (Grant No. 17PJ1406200). L. X. Y. acknowledges the support from Tsinghua University Initiative Scientific Research Program. Y. W. L. acknowledges the support from China Scholarship Council.